\documentclass{article}

\usepackage{arxiv}

\usepackage[utf8]{inputenc} 
\usepackage[T1]{fontenc}    
\usepackage{hyperref}       
\usepackage{url}            
\usepackage{booktabs}       
\usepackage{amsfonts}       
\usepackage{nicefrac}       
\usepackage{microtype}      
\usepackage{lipsum}
\usepackage{graphicx}
\graphicspath{ {./images/} }
\usepackage{lmodern} 
\usepackage{microtype}
\usepackage[numbers,square,sort&compress]{natbib}
\usepackage[toc,page]{appendix} 
\usepackage{lipsum, amsmath}

\title{Introducing Discipline Score Based on League Overall Swinging Probability}

\author{
 Wuhuan Deng \\
  Department of Applied Mathematics\\
  University of Washington\\
  Seattle, WA 98105 \\
  \texttt{wudeng@uw.edu} \\
   \And
 Scott Nestler \\
  Department of Sports Management\\
  University of Florida\\
  Gainesville, FL 32612 \\
  \texttt{nestler.scott@ufl.edu} \\
}

\begin{document}
\maketitle
	
	\begin{abstract}
		Plate discipline is an important feature of a hitter’s success. Hitters who are able to recognize good pitches to swing at and balls to take are generally recognized as disciplined hitters. Although there are some metrics that can provide insight into the patience of a hitter, most do not capture the ability of a batter to take balls. In this research, we introduce two new metrics, Discipline Score (DS) and Adjusted Discipline Score(ADS), which evaluate batters' discipline when the pitch is a ball compared with the predicted tendencies of all batters in the league.
	\end{abstract}

	
	\section{Introduction}
	\subsection{Motivation}
	Plate discipline is one of the key components of consistent offense. Generally, a hitter with good plate discipline is able to generate more walks and fewer strikeouts. Vock and Vock found that improved plate discipline increases batting average, on-base percentage, and slugging percentage \cite{vv}. When a batter gets a walk, it not only means that he is on base, but it also causes the pitcher to use at least 4 (of a limited number of) pitches. When we look at Major League Baseball (MLB) today, players who can generate many walks are all big stars in the league. Therefore, both the front office and the coaches should be able to recognize players who have great discipline, and this could help them make personnel decisions during free agency and trades. The 5 players who walked the most in 2024 are shown in Table~\ref{top5walkers}.
	
	\begin{table}[h]
		\centering
		\begin{tabular}{c c}
            \hline
			Player & Number of Walks \\
			\hline
			Aaron Judge & 133 \\
			Juan Soto & 129 \\
			Kyle Schwarber & 106 \\
			Shohei Ohtani & 81 \\
			Ian Happ & 80 \\
            \hline
		\end{tabular}
		\caption{5 players who generated most walks in 2024 season.}
        \label{top5walkers}
	\end{table}

    On top of that, Run expectancy (RE) models describe how many runs a team is expected to score given the current game state of outs and baserunners. Recent work~\cite{RE} represents RE as a Markov process, where each at-bat changes the game state through probabilistic transitions with runs as rewards. Since plate discipline affects this transition, it directly shapes the expected run output of a team. This is also why we need metrics to identify disciplined hitters.

	\subsection{Goal}
	Our goal of this research is to evaluate the discipline of a batter compared to the entire league when the pitch is a ball. The probability of the entire league swinging on a given ball will be predicted, and then the Discipline Score (DS) of the batter on this ball will be determined based on the actual result of this pitch. Instead of representing how many walks a batter can generate in a season, this method provides a deeper insight when a batter faces a ball. And we can also know the discipline of batters when facing different types of pitches, such as fastball, breaking ball and offspeed, by adjusting the input pitch data.
	
	\section{Related Works}
\begin{figure}
        \centering
        \includegraphics[width=\textwidth]{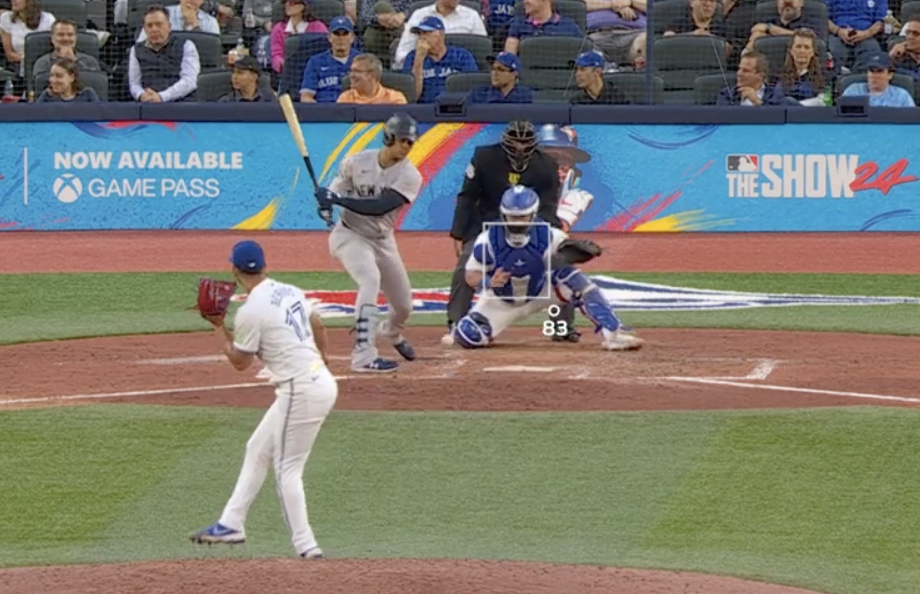}
        \caption{Juan Soto swung on a ball.}
        \label{soto}
    \end{figure}

\begin{figure}
        \centering
        \includegraphics[width=\textwidth]{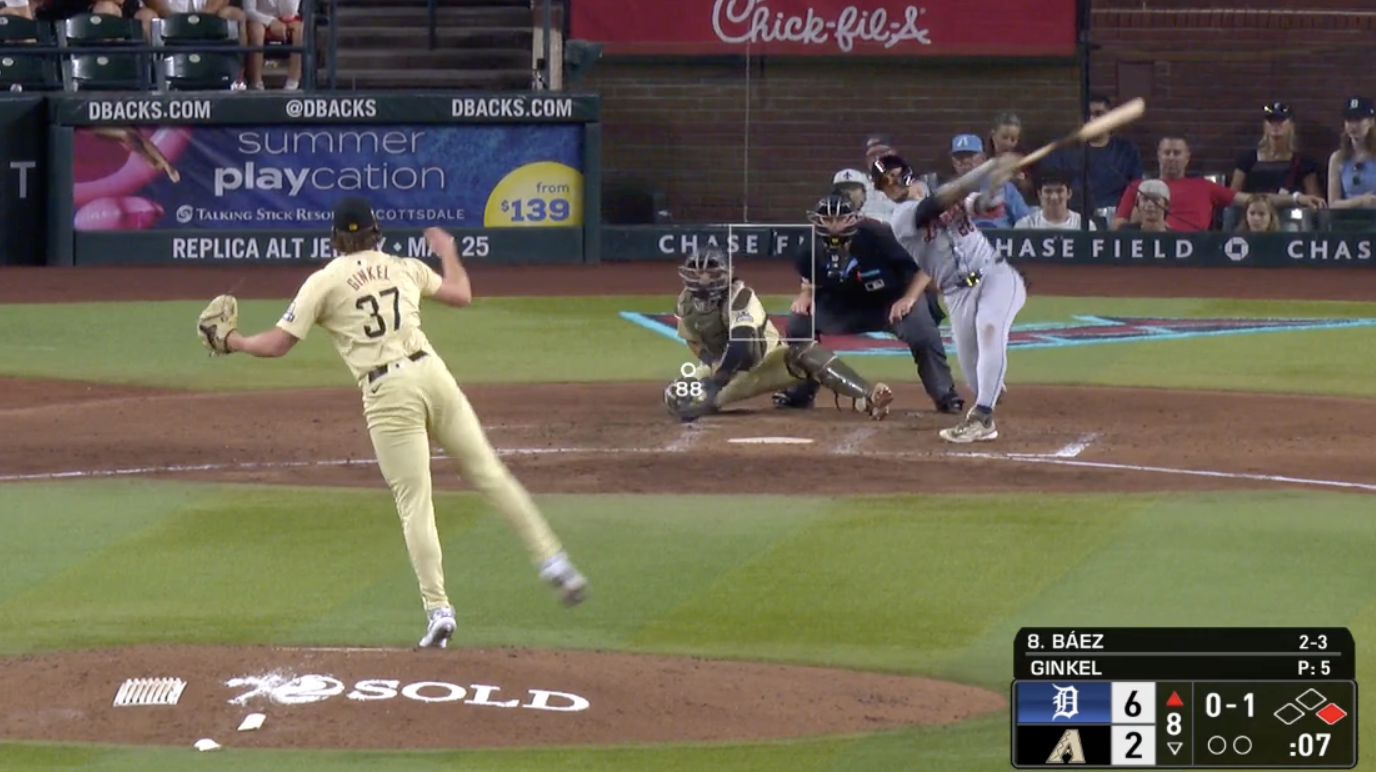}
        \caption{Javier Baez swung on a ball.}
        \label{baez}
    \end{figure}

	Traditional metrics, such as O-Swing\%, which evaluates how often a batter swings at pitches outside the strike zone and Z-Swing\%,  which measures how often a batter swings at pitches inside the strike zone have their limitations \cite{tra}. O-Swing\% does not consider the differences between each pitch. If two batters face 10 pitches outside strike zone, and they both swing 3 times, they will get same O-Swing\%, 0.3. However, the 10 pitches that they face can be quite different -- one may swing at an outside pitch which is very close to the strike zone and another one may swing at an outside pitch which is far from the strike zone. But this difference is not reflected by O-Swing\%. In Figure~\ref{soto} and ~\ref{baez}, although both pitches are balls and the batters have swung, it is obvious that those two pitches are different.  Javier Baez obviously swung on a ball which should be identified as such by a big league player. However, Juan Soto swung at a ball that was not too far from the edge of the strike zone.

	In 2023, Ryan Yee and Sameer K. Deshpande introduced a three-step framework to evaluate plate discipline. In their first step, they fitted a flexible Bayesian nonparametric model to predict (i) the probability that a batter makes contact; (ii) the probability that the umpire calls a strike; and (iii) the average number of runs that the batting team is expected to score \cite{ba}. There are two more steps in their research, but only the first step is directly related to this study. When they predicted the contact probability, the model only considers pitches at which batters swung. And they evaluated plate discipline based on the result of contact and expected runs. They also predicted the probability that the umpire calls a strike, but in this research we used the exact definition of a ball based on pitch location data. MLB’s EAGLE (Expected Aggressiveness, Location, and Expectancy) model estimates the run-value impact of a hitter’s swing decision by comparing actual swing choices to league-average expected values given pitch location and count \cite{EAGLE}.

	\section{Define Discipline Score \& Adjusted Discipline Score}
	\subsection{Discipline Score}
	Discipline Score (DS) evaluates a batter's discipline based on the overall tendency of players in the league to swing at a given ball. For a given ball, DS is defined as:
	\begin{equation*}
		DS = (-1)^R \cdot (R (1 - P_S) + (1 - R) P_S),
	\end{equation*}
	where $ R $ is the actual decision of the batter ($R = 1$ for swing or $R = 0$ for take), and $ P_S $ is the probability that the entire league will swing at the given ball. DS ranges from -1 to 1 with larger values indicating better discipline. $ (-1)^R $ determines the sign of DS, where a negative value suggests that the batter swings at the ball. and a positive value means that the batter takes the ball. Table~\ref{DScalc} shows several examples of possible DS values.
	
    \begin{table}[h]
        \centering
        \small
		\begin{tabular}{c c c c}
			\hline
			$ P_S $ & $ R $ & $ DS $ & Interpretation \\
			\hline
			0.7 & 0 & 0.7 & Take a ball that most batters swing at \\
			0.7 & 1 & -0.3 & Swing at a ball that most batters swing at \\
			0.3 & 0 & 0.3 & Take a ball that most batters take \\
			0.3 & 1 & -0.7 & Swing at a ball that most batters take \\
			\hline
		\end{tabular}
		\caption{Example values of DS.}
            \label{DScalc}
	\end{table}

	\subsection{Adjusted Discipline Score}
	DS only considers the action of swing or take itself, but ignores the result if the batter swings at the ball. Sometimes, a batter can generate a home run even swinging at a ball. Therefore, we introduced a new metric Adjusted Discipline Score (ADS), which for a given pitch is defined as:
	\begin{equation*}
		ADS = DS + CQ,
	\end{equation*}
	where CQ is the contact quality which is defined as:
	\[
	CQ = 
	\begin{cases}
		EV_{score} \times LA_{score}, \text{ if batter makes contact and not foul} \\
		0, \text{ otherwise (including swing and miss)}
	\end{cases}
	\]
	where $ EV_{score} = min(1, max(0, \frac{EV - 70}{28})) $ and $ LA_{score} = max(0, 1 - \frac{|LA - 20|}{20}) $. EV is the exit velocity and LA is the launch angle of the given pitch. By definition, 98mph is the smallest EV for a barrel ~\cite{barrel}, so when EV is larger than 98mph, $ EV_{score} $ will be 1. And when EV is between 70mph and 98mpb, $ EV_{score} $ will be between 0 and 1. When EV is smaller than 70, which will be considered as weak contact, $ EV_{score} $ will be 0 so that the value of $ CQ $ will also be 0. Launch Angle between 10 and 30 degrees is considered as a hard hit range. When the value of $ LA $ is smaller than 0 or larger than 40 degrees, $ LA_{score} $ will also be 0. And the value of $ LA_{score} $ is maximized at 1 when $ LA = 20 $. Since the smallest value of CQ is 0, ADS is definitely larger or equal to DS.

    \begin{table}[h]
        \centering
        \small
        \begin{tabular}{c c c c c}
            \hline
            $ EV $ & $ EV_{score} $ & $ LA $ & $ LA_{score} $ & CQ \\
            \hline
            80 & 0.357 & 20 & 1 & 0.357 \\
            100 & 1 & 10 & 0.5 & 0.5 \\
            60 & 0 & -5 & 0 & 0 \\
            80 & 0.357 & 45 & 0 & 0\\ 
            \hline
        \end{tabular}
        \caption{Example values of $ EV, LA, EV_{score}, LA_{score}, CQ $}
        \label{tab:placeholder}
    \end{table}

    \section{Estimate the League Overall Swinging Probability}
	\subsection{Dataset}
	Statcast data are captured through MLB’s Hawk-Eye optical tracking system, which uses a network of twelve high-speed cameras operating at up to 300 frames per second to record the full 3D trajectory, spin axis, and seam orientation of each pitch with sub-inch precision~\cite{Statcast}. This level of measurement accuracy ensures that our pitch-level features (e.g., location, velocity, and spin) are grounded in highly reliable, physics-based observations rather than modeled estimates.

    In this study, we use pitch-by-pitch tracking data from the MLB Statcast database. It can be accessed using the \textit{pybaseball} library in \textit{python} or the \textit{baseballr} package in \textit{R}. In this research we use data from the 2021 - 2023 regular seasons for the model training process, and the 2024 regular season for player analysis. Due to factors such as the Statcast camera being obstructed during a pitch, a very small number of pitches were not recorded. The dataset of 2021 - 2023 regular seasons contains 1,773,300 pitches, and 1,766,439 pitches (or 99.6\%) are successfully tracked. We do not determine a pitch being a ball based on umpire's call. Instead we define a pitch as a ball when it has has no intersection with the strike-zone. Under this definition, there are 941,407 tracked balls from 2021 - 2023 and 370,489 tracked balls from 2024 season. We selected the 6 features shown in Table~\ref{features}, including pitch location data, pitch characteristics, and game context for training the model and the batter's actual result (swing or take the pitch) as label.

	\begin{table}
		\centering
        \small
		\begin{tabular}{c c}
			\hline
			Feature Name & Description \\
			\hline
			plate\_x & Horizontal location of a pitch at home plate \\
			plate\_z & Vertical location of a pitch at home plate \\
			release\_speed & Velocity of a pitch when it leaves pitcher's hand \\
			release\_spin\_rate & Number of times the ball spins when it leave pitcher's hand \\
			pfx\_x & Horizontal deviation in feet \\
			pfx\_z & Vertical deviation in feet \\
			\hline
		\end{tabular}
		\caption{This table shows the 6 input features name and brief description.}
            \label{features}
	\end{table}
	
	Since the top and bottom of strike zone vary based on each batter's stance -- the top is a horizontal line at the midpoint between the top of the batter's shoulder and the top of the uniform pants, and the bottom is a horizontal line at the hollow beneath the kneecap -- we normalize vertical position before estimating league swinging probability:
	\begin{equation*}
		norm\_plate\_z = \frac{plate\_z - sz\_bot}{sz\_top - sz\_bot},
	\end{equation*}
	where $ plate\_z $ is the vertical location of a pitch at home plate, $ sz\_bot $ is the vertical position of the bottom line of the strike zone, and $ sz\_top $ is the vertical position of the top line of the strike zone. In this way, we removed the effect of each batter's size on the strike-zone dimensions, and it will be easier to compare between different batters.

    To ensure that the estimated swinging probability is reasonable, pitches are divided into three categories: Fastball, Breaking Ball, and Offspeed.

	\subsection{Estimate $ P_S $}
    After cleaning and preparing the data, we implement a non-parametric method to estimate {$P_S$} based on the similarity of new pitches given the feature space, e.g. location, velocity, etc. Given a new pitch from the 2024 regular season, we find the $ k $ most similar pitches from the 2021 to 2023 seasons and compute the empirical swinging probability among those pitches. This method is inspired by the k-nearest neighbors (K-NN) method, which has previously been shown to perform well in baseball applications. 
    For instance, Ross et al. used K-NN to model the relationship between launch angle, exit velocity, and expected wOBA when optimizing hitter performance, demonstrating its effectiveness for estimating outcomes from comparable batted-ball profiles ~\cite{ross2018launchangle}.

    Let $ x_i \in R^d $ be the feature vector of the $ i $-th pitch in the training set (2021 - 2023 regular season dataset), and $ d $ be the number of features. Given a new pitch $ \hat{x} $, the estimated swinging probability is defined as:
    $$ P_S (\hat{x}) = \frac{1}{k} \Sigma_{j \in N(\hat{x})} R_j, $$
    where $ N(\hat{x}) $ is the set of $ k $ nearest training pitches to $ \hat{x} $ in Euclidean distance. In our research, we test four values of $ k $: 10, 100, 200, and 500.

    To evaluate the reliability and probabilistic accuracy of our swing–decision model, we employ both the Brier score and the calibration curve. 
    The Brier score~\cite{BrierScore} is a proper scoring rule that quantifies the mean squared difference between the predicted probability $\hat{p}_i$ and the actual binary outcome $y_i \in \{0,1\}$, defined as:
    \[
    \text{Brier Score} = \frac{1}{N} \sum_{i=1}^{N} (y_i - \hat{p}_i)^2.
    \]
    Lower values indicate better overall performance, capturing both discrimination and calibration of probabilistic predictions.  
    
    The \textit{calibration curve} provides a visual method of probabilistic accuracy. Predicted probabilities are grouped into bins (e.g., 0.0–0.1, 0.1–0.2, etc.), and the average predicted probability within each bin is plotted against the actual proportion of observed positive outcomes in that same bin. A perfectly calibrated model will produce points lying close to the 45° line.

\begin{figure}
        \centering
        \includegraphics[width=\textwidth]{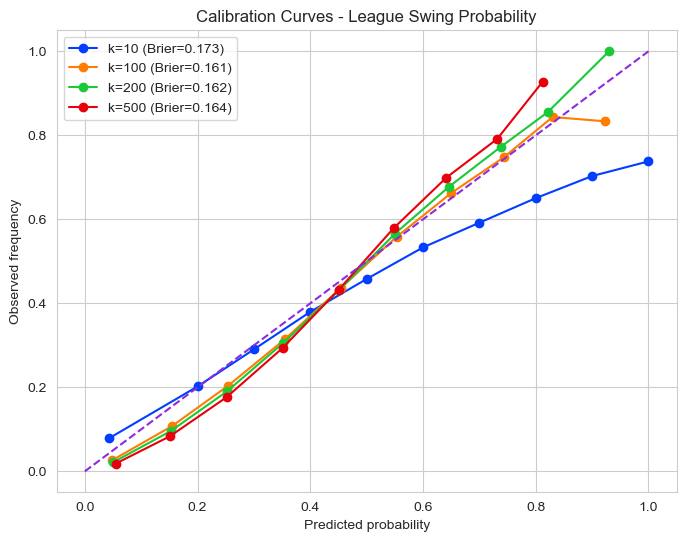}
        \caption{Calibration curve and brier score of each value of $ k $.}
        \label{fig:ktest}
    \end{figure}

    We input each value $ k $ and compute the Brier score and plot the calibration curve for 2024 regular season data. In Figure~\ref{fig:ktest}, $ k = 100 $ and $ k = 200 $ have very similar Brier scores, 0.161 and 0.162, which are lower than the other two. However, $ k = 100 $ cannot catch the tendency when the swinging probability is higher than 0.8. When $ k = 200 $, the calibration curve is more stable and smooth compared with $ k = 100 $. Therefore, in the following analysis, we choose $ k = 200 $ for the model.

    \begin{figure}
        \centering
        \includegraphics[width=\textwidth]{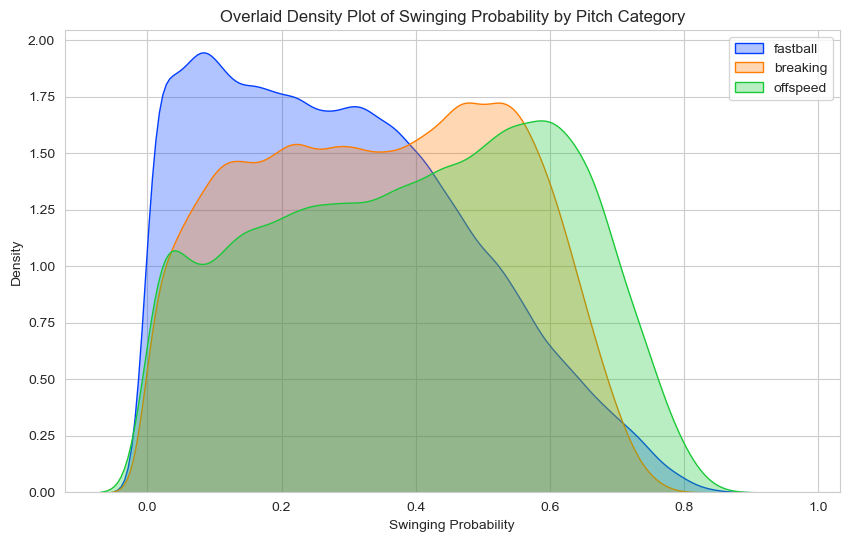}
        \caption{League swinging probability for each pitch category of 2024 regular season.}
        \label{fig:LeagueSwingProb}
    \end{figure}

    In Figure~\ref{fig:LeagueSwingProb}, we plot the distribution of estimated swinging probability for pitches identified as balls for each pitch category. The value of Y-axis is density, so it allows us to compare across different pitch categories regardless of sample size. Fastballs (blue) shows a strong left-skewed distribution, with most of $ P_S $ is less than 0.4. This suggests that the entire league is not fooled by fastballs easily. This is reasonable because of more predictable trajectories of fastballs. Breaking balls (red) have a more balanced distribution, with highest density around $ P_S = 0.5 $. This indicates that compared with fastballs, breaking balls are more likely to lead a swing when they are actually outside the strikezone. Offspeed pitches (green) have a large portion in the higher probabilities region ($ > 0.5 $). This makes sense since offspeed pitches are more deceptive.

    \begin{table}
        \centering
        \small
        \begin{tabular}{@{}l c l c l c@{}}
        \toprule
        \multicolumn{2}{c}{\textbf{Fastball}} &
        \multicolumn{2}{c}{\textbf{Breaking Ball}} &
        \multicolumn{2}{c}{\textbf{Offspeed}} \\
        \midrule
        Pitcher & $P_S$ & Pitcher & $P_S$ & Pitcher & $P_S$ \\
        \midrule
        Jose Soriano    & 0.357 & Aaron Nola    & 0.443 & Logan Webb      & 0.545 \\
        Zack Wheeler    & 0.356 & Luke Jackson  & 0.437 & Cooper Criswell & 0.500 \\
        Kutter Crawford & 0.354 & Logan Gilbert & 0.421 & Paul Skenes     & 0.486 \\
        Garrett Crochet & 0.348 & Mitch Spence  & 0.410 & Zack Littell    & 0.486 \\
        Dylan Cease     & 0.348 & Bryan Abreu   & 0.404 & Luis Castillo   & 0.472 \\
        \bottomrule
        \end{tabular}
        \caption{5 highest $ P_S $ value among pitchers of each pitch category with qualified number of pitches thrown in 2024 regular season.}
    \end{table}

    \section{Results}
    We compute the DS and ADS of each batter who played in 2024 by using the league overall swinging probability from the 2021-2023 seasons. After computing DS and ADS value of each pitched ball, we compute the average value of DS and ADS of each batter for the entire regular season to have a general understanding of entire season.

    \begin{table}
        \centering
        \small
        \begin{tabular}{@{}l c l c l c@{}}
        \toprule
        \multicolumn{2}{c}{\textbf{Fastball}} &
        \multicolumn{2}{c}{\textbf{Breaking Ball}} &
        \multicolumn{2}{c}{\textbf{Offspeed}} \\
        \midrule
        Batter & $DS$ & Batter & $DS$ & Batter & $DS$ \\
        \midrule
        Jesse Winker     & 0.146 & Juan Soto        & 0.199 & Jake Bauers     & 0.239 \\
        Jonathan India   & 0.139 & Dylan Moore      & 0.175 & Juan Soto       & 0.223 \\
        J.P. Crawford    & 0.130 & Andrew McCutchen & 0.174 & Josh Rojas      & 0.187 \\
        Yandy Díaz       & 0.116 & Ha-Seong Kim     & 0.170 & Lars Nootbaar   & 0.176 \\
        Ian Happ         & 0.115 & Aaron Judge      & 0.170 & Aaron Judge     & 0.175 \\
        \bottomrule
        \end{tabular}
        \caption{Table of the highest 5 $ DS $ value of each pitch category in 2024 regular season.}
        \label{5highestDS}
    \end{table}

    \begin{figure}
        \centering
        \includegraphics[width=\textwidth]{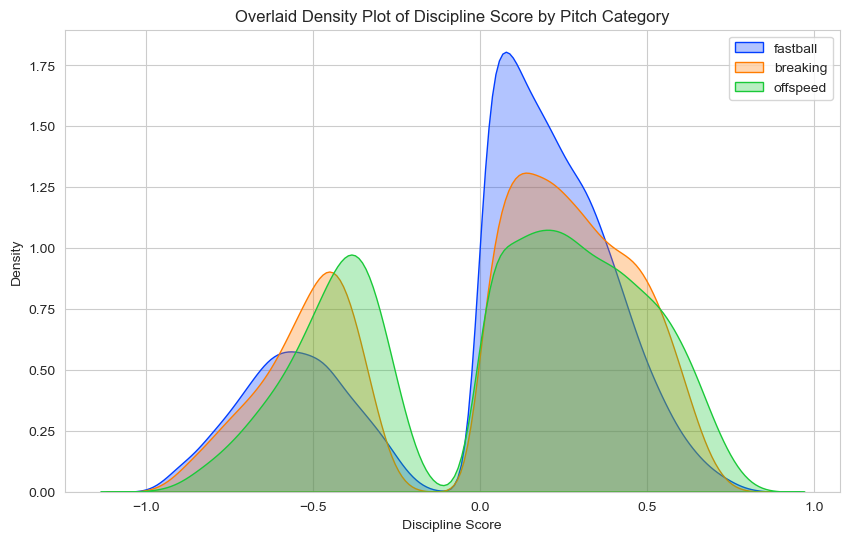}
        \caption{Distribution of $ DS $ value of every pitch among all categories.}
        \label{DSdist}
    \end{figure}

    \subsection{Compare Players with Similar Stats in 2024}
\begin{figure}
        \centering
        \includegraphics[width=\textwidth]{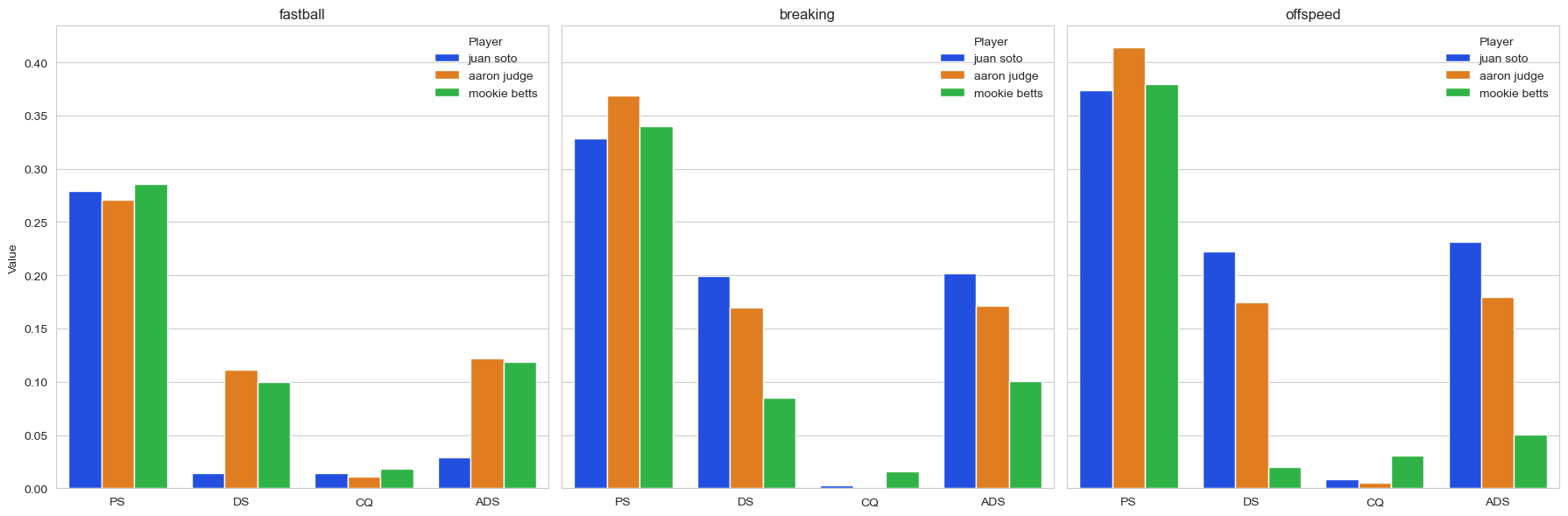}
        \caption{Histograms of $ PS, DS, CQ, ADS $ respect to each pitch category and all together for Aaron Judge, Juan Soto and Mookie Betts in 2024 season.}
        \label{JudgeSotoBetts}
    \end{figure}

    In Figure~\ref{JudgeSotoBetts}, we compare Juan Soto, who is regarded as one of the most discipline player in the league, with other two outstanding players Aaron Judge and Mookie Betts. Aaron Judge and Juan Soto have exactly same O-Swing\% and similar BB\%, but Aaron Judge has a much higher K\% than Juan Soto in 2024. However, Aaron Judge obtains a higher $ DS $ value among all categories pitches, which may be caused by a very high $ DS $ in fastball. In comparison, Juan Soto performs better when facing breaking and offspeed pitches. Mookie Betts has a lower BB/K than Juan Soto with a lower BB\% but also a lower K\%. Again, Juan Soto performs worse when facing fastball, but he does an outstanding job when facing breaking and offspeed pitches.

    \subsection{Compare DS with other Metrics}
    \begin{figure}
        \centering
        \includegraphics[width=\textwidth]{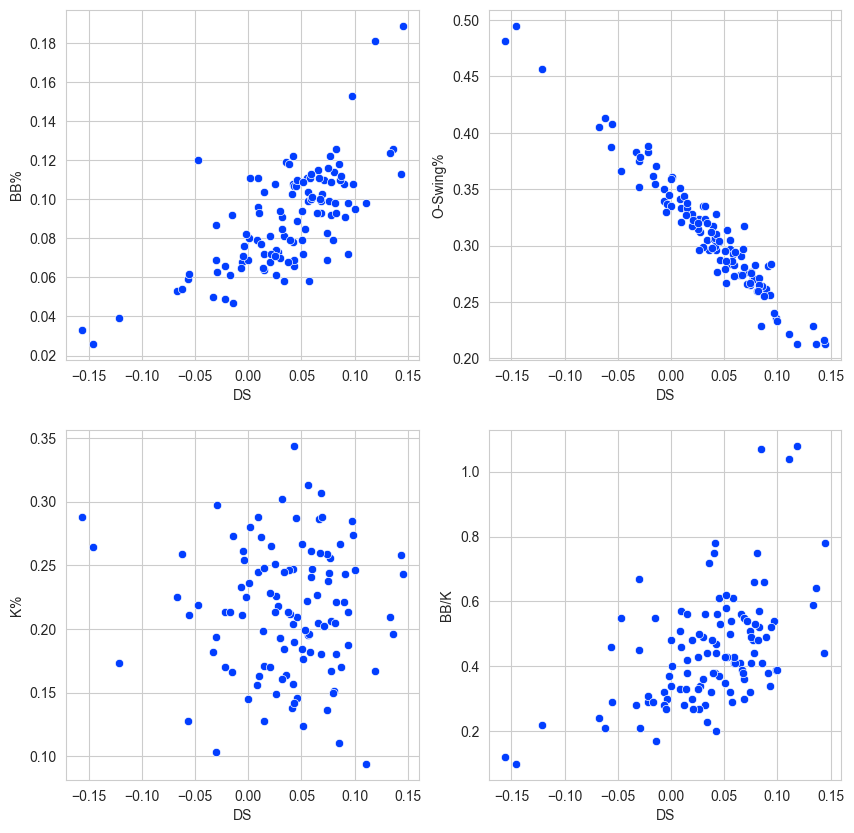}
        \caption{Scatter plots of relationship between DS and BB\%, O-Swing\%, K\%, K/BB.}
        \label{DSscatter}
    \end{figure}

    In Figure~\ref{DSscatter}, we plot the relationship between $ DS $ and other discipline-related metrics value of entire league in 2024 season and players who had at least 150 Plate Appearances are considered in this research. It is obvious that $ DS $ and BB\% have a strong positive connection, and although BB/K and $ DS $ also have a positive connection, the connection is weaker compared with BB\%. $ DS $ clearly has a very strong negative relation with O-Swing\%, which is expected, since O-Swing\% measures how often a batter swings at a ball, and this is highly related to discipline. However, there is no clear relation between K\% and $ DS $. This is reasonable, since if batters take a strike inside the zone, they will also be more likely to strikeout and $ DS $ does not deal with any pitches inside the strike zone.

    \section{Conclusion}
    This research introduces the Discipline Score (DS) and the the Adjusted Discipline Score (ADS) as innovative, context-sensitive metrics to evaluate plate discipline, moving beyond existing measures that treat all pitches equivalently. By leveraging league-wide swinging probabilities for each ball, DS provides a granular understanding of a batter's decision-making, reflecting not just outcomes at the plate but the quality of choices in relation to league norms. Coupled with ADS, which incorporates Contact Quality, these metrics offer a more comprehensive framework for distinguishing truly disciplined hitters, those able to maximize value from tough pitching decisions.

    Our findings highlight strong relationships between DS and key offensive metrics such as BB\% and O-Swing\%, demonstrating the validity and practical utility of the proposed approach. The DS framework enables coaches, analysts, and front offices to better identify undervalued player skills, tailor training interventions, and inform personnel strategies in areas such as scouting and contract decisions. Future research may extend this methodology to incorporate deeper pitch characteristics, batter situational context, and predictive analytics for player development and performance forecasting. In general, DS and ADS lay the groundwork for more precise and data-driven evaluation of plate discipline, potentially transforming quantitative analysis in baseball.
    

\end{document}